\documentclass[twocolumn,superscriptaddress,showpacs,floatfix]{revtex4}
\usepackage{graphicx}

\usepackage{bm}
\usepackage{amsmath, amssymb, color}

\def\a{\alpha}

\def\om{\omega}

\def\bC{{\bf C}}
\def\bCd{{\bf C}^\dagger}

\definecolor{grey}{rgb}{0.4,0.4,0.5}
\definecolor{darkgreen}{rgb}{0,0.5,0}
\definecolor{darkred}{rgb}{0.6,0.0,0}
\definecolor{lightbrown}{rgb}{1,0.9,0.8}
\definecolor{brown}{rgb}{0.6,0.3,0.3}
\definecolor{darkblue}{rgb}{0,0,0.8}
\definecolor{darkmagenta}{rgb}{0.5,0,0.5}

\def\({\left(}
\def\){\right)}

\def\la{\label}

\def\be{\begin{equation}}
\def\ee{\end{equation}}
\def\ben{\begin{equation*}}
\def\een{\end{equation*}}

\def\J{J(\alpha)}
\def\z{z(\alpha)}
\def\K{\hat{K}}
\newcommand{\avg}[1]{\left< #1 \right>}
\newcommand{\erf}[1]{{\rm erf}\left(#1\right)}
\newcommand{\erfc}[1]{{\rm erfc}\left(#1\right)}

\newcommand{\en}{\end{equation}}
\newcommand{\tpsi}{\tilde{\psi}}

\newcommand{\bea}{\be\begin{aligned}}
\newcommand{\eea}{\end{aligned}\ee}
\newcommand{\bean}{\ben\begin{aligned}}
\newcommand{\eean}{\end{aligned}\een}

\newcommand{\bei}{\begin{itemize}}
\newcommand{\eei}{\end{itemize}}

\newcommand{\bee}{\begin{enumerate}}
\newcommand{\eee}{\end{enumerate}}

\newcommand{\bem}{\left (\begin{matrix}}
\newcommand{\eem}{\end{matrix} \right )}

\def\bc{c}
\def\bcd{c^\dagger}
\def\bn{n}

\def\bC{C}
\def\bCd{C^\dagger}

\begin{document}
 
\title{Scaling of critical wavefunctions at topological Anderson transitions in 1D }

\author{Eoin Quinn}
\affiliation{Max-Planck-Institut f\"ur Physik komplexer Systeme, N\"othnitzer Str.\ 38, 01187
  Dresden, Germany} 
\author{Thomas Cope}
\affiliation{School of Mathematical Sciences, University of Nottingham, Nottingham NG7 2RD, United Kingdom}
\author{Jens H. Bardarson}
\affiliation{Max-Planck-Institut f\"ur Physik komplexer Systeme, N\"othnitzer Str.\ 38, 01187
  Dresden, Germany} 
\author{Alexander Ossipov}
\affiliation{School of Mathematical Sciences, University of Nottingham, Nottingham NG7 2RD, United Kingdom}

\date{\today}

\begin{abstract}

  Topological Anderson transitions, which are direct phase transitions between topologically distinct Anderson localised phases, allow for criticality in 1D disordered systems. We analyse the statistical properties of  an emsemble of critical wavefunctions at such transitions. We find that the local moments are strongly inhomogeneous, with significant amplification towards the edges of the system. In particular, we obtain an analytic expression for the spatial profile of the  local moments which is valid at all topological Anderson transitions in 1D, as we verify by direct comparison with numerical simulations of various lattice models.
    
  \end{abstract}


\maketitle 

%\tableofcontents

%\section{Introduction}

The interplay of topology and disorder in electronic systems can give rise to critical transitions between localised phases \cite{BMSA98,BCF00,BFG00,GRV05,MSHP13,ABK15}.
This is of particular interest in 1D, as here there are good reasons to expect a system of  electrons to realise an insulating phase, even in the absence of  electronic correlations. 
On the one hand, a 1D chain is generically unstable to dimerisation via Peierls distortion,
while on the other, any amount of extensive disorder leads to Anderson localisation. Remarkably however, these two effects can be tuned in combination to induce critical behaviour. 
This occurs as distinct dimerisation patterns realise distinct topological phases, whereas strong disorder washes out the detail of the dimerisation, which drives the system between the phases. 
In general, such critical transitions occur in 1D for disordered systems which realise symmetry-protected topological phases, and are described by two-parameter scaling \cite{ABK15}.
We shall refer to the transitions as topological Anderson transitions, and it is the resulting critical states that are of interest to us in this paper.

Whether a system may realise distinct topological phases is dictated by both its dimension and  the anti-unitary symmetries of its Hamiltonian \cite{Schnyder2008,Kitaev2009,Ryu2010}. In particular, for each dimension a certain five of the ten  symmetry classes  of the Altland-Zirnbauer classification \cite{AZ} allow for non-trivial topological phases. In 1D these are the three chiral classes (AIII, BDI, CII) and two of the Bogoliubov-de Gennes classes (D, DIII).  The situation in 1D is however more special as here, irrespective of symmetry class, all topological Anderson transitions share the same critical properties, a phenomenon dubbed \emph{superuniversality} \cite{GRV05}.  This was established by identifying exact mappings between representative single-channel models for each of the five classes, and by showing that for a long chain the properties of a system with many channels are determined by a dominant effective single channel.

It is of great interest to understand the properties of the critical wavefuntion at  this superuniversal transition. This has been partially investigated by Balents and Fisher \cite{BF97} who calculated the system-size scaling of the moments of the normalised wavefunction in the bulk and their spatial correlations
\be\la{BF}
\langle |\psi(x)|^{2q}\rangle \propto \frac{1}{N} \,,\quad \langle |\psi(x_1)\psi(x_2)|^{q}\rangle \propto  \frac{1}{N} |x_1 - x_2|^{-\frac{3}{2}}\,,
\ee
where $\langle \ldots\rangle$ denotes an averaging over a disorder ensemble, $N$ is the system size and $q>0$.  The  $1/N$-scaling of the moments is characteristic of localisation, while the algebraic decay of the spatial correlations reflects the critical nature of the wavefunction \cite{EM08}.

Here we analytically show that the local moments are in fact inhomogeneous throughout the system. In the bulk they scale as 
\be\la{inhom}
\langle |\psi(x)|^{2q}\rangle \propto \frac{1}{\sqrt{x(N-x)}}\,,
\ee
and directly at the edges they scale as $1/\sqrt{N}$.  Equation (\ref{inhom}) describes how the $1/N$-scaling in the bulk changes to  $1/\sqrt{N}$ towards the edges.  The strong  amplification of the disorder-averaged wavefunction  near the boundaries $\langle |\psi(1)|^{2q}\rangle/\langle |\psi(N/2)|^{2q}\rangle\propto \sqrt{N}$  distinguishes it significantly from all other critical wavefunctions, and can be understood as a remnant of the topological edge mode that is created/destroyed by the transition.

These results are derived from a specific single-channel model in class BDI for which the critical wavefunction at the transition is exact and analytically tractable. Based on superuniversality, we argue that the results are generally valid at all 1D  topological Anderson transitions. We verify this by directly comparing them with numerics for both a class BDI model with no $E=0$  wavefunctions away from the $N\to\infty$ limit, and for a multi-channel model belonging to symmetry class D.

%%%%%%%%%%%%%%%%% Section %%%%%%%%%%%%%%%%%%%%%%%%%%
%\section{Exact critical wavefunction  }   
%%%%%%%%%%%%%%%%% Section %%%%%%%%%%%%%%%%%%%%%%%%%%
\medskip

To explore the critical physics of 1D topological Anderson transitions we focus on a disordered dimerized tight-binding chain.
This model appears as the static limit of the Su-Schrieffer-Heeger model \cite{SSH, SuSchriefferHeegerRMP}, which is used for example to describe polymers such as polyacetylene. 
For reasons that will be made clear shortly, we primarily focus on a chain of odd length $2N+1$, and label the sites from 0 to $2N$:
\be\la{poly}
H = \sum_{j=1}^{N}\big[-t_{j} \bcd_{2 j-2} \bc_{2j-1}-t'_j \bcd_{2j-1} \bc_{2j}+{\rm h.c.}\big]\,.
\ee
Disorder appears through the hopping amplitudes
\be
t_j = t + W \omega_j\,,\quad t'_j = t' + W' \omega'_j\,,
\ee
where $\om_j$, $\om'_j$ are independent random variables uniformly distributed on the interval $[-0.5,0.5]$. The model is invariant under the time reversal symmetry $T=K$, where $K$ is complex conjugation, and has a chiral symmetry $SH S^{-1}=-H$, with $S= \sum_{j=0}^{2N} (-1)^j \bn_j$. The parity operator $P =TS$ squares to the identity when acting on single-particle states, and so the model belongs to the BDI symmetry class.

The odd length of the chain guarantees the existence of a zero-energy eigenstate for any set of parameters $t_j$, $t'_j$ \cite{FL77,BRFHMM02}. This follows from the chiral symmetry of the model which restricts the spectrum to be symmetric about zero. The Schr\"odinger equation at zero energy reduces to a simple recursion relation with solution
\be\la{wf}
\tpsi_{2j} = \left(\prod_{k=1}^j \frac{t_{k}}{t'_{k}} \right)\tpsi_0\,,\quad \tpsi_{2j-1} =0 \,,\quad j=1,\ldots,N\,, 
\ee
where the $\sim$ indicates that this wavefunction is unnormalised.
This eigenstate describes a topological boundary mode that is 
%exponentially 
localised at one edge of the chain. 
For the clean case ($t_k=t$, $t'_k=t'$) the mode resides at the left edge if $|t|<|t'|$ and at the right edge if $|t|>|t'|$, while precisely at $|t|=|t'|$  there is a topological transition at which the  wavefunction is uniform. In the presence of disorder this transition becomes a topological Anderson transition, and it can be driven purely by a change in the disorder strength.

It is worthwhile to contrast this behaviour of the edge mode with that for an even length chain, for which there are boundary modes at both edges of the chain on one side of the transition, and no boundary modes on the other side. The connection can be made by considering adding an extra site to the odd length chain. On its own the extra site has a single localised zero energy mode. On coupling it to an end of the chain this localised mode will either remain (protected by the chiral symmetry) if there is no zero energy state at that edge, or it will  combine with and annihilate the boundary mode if there is one present. 
More formally, for the even length chain the two sides of the transition can be distinguished by the Zak-Berry phase \cite{Berry, Zak}, which is a ${\mathbb Z}_2$ bulk invariant, and the extra site (which breaks the 2-site unit cell) can be thought of as giving rise to a defect mode \cite{JRL14}. We remark that for the odd length chain there are two choices for the unit cell, and so there is no natural way to call one phase trivial and the other topological.
%but nevertheless the Zak-Berry phase will jump across the transition.

%%%%%%%%%%%%%%%%% Section %%%%%%%%%%%%%%%%%%%%%%%%%%
%\section{Analytic}
%%%%%%%%%%%%%%%%% Section %%%%%%%%%%%%%%%%%%%%%%%%%%
\medskip

We  now calculate the local moments per unit cell of the normalized zero-energy wavefunction:
\be
I_q(n)=\Big\langle\sum_{j\in\Lambda_n}\big|\psi_{j}\big|^{2q}\Big\rangle\,,
\ee 
where $n$ indexes the unit cells of the underlying lattice, and $\Lambda_n$ is the set of sites in the $n^{\rm th}$ unit cell.
In the present case  $I_q(n)=\avg{|\psi_{2n}|^{2q}}$, with $\psi_{2n}\equiv \tpsi_{2n}/(\sum_{m=0}^N|\tpsi_{2m}|^2)^{1/2}$, as the wavefunction has support on only the even sites of the chain.
The normalisation factor contains non-trivial dependence on the disorder parameters $\omega_k$, $\omega'_k$, and to handle this we use a technique developed for continuum models that maps the calculation to one in Liouville quantum mechanics \cite{ST98,BF97}. In the present case, working explicitly with a lattice model allows for a controlled description of the problem in the strong disorder regime.
We outline here only the main steps of the calculation and defer the details  to the Appendix. 

The disorder averaged local moments of the wavefunction $I_q(n)$ can be expressed as a functional integral over random walk configurations of the variables $ z_k=2\ln|t_k/t'_k|$. In the thermodynamic limit $N\to\infty$, the behavior of this integral is determined by the first two moments of the distribution function of the $z_k$. As a result, evaluating  $I_q(n)$ reduces to solving a discrete-time version of Liouville quantum mechanics -- the quantum mechanics of a particle moving in an exponential potential. This latter problem can be solved  by approximating the exponential potential by a hard wall. 

As shown in the Appendix, the averaged local moments can be expressed in terms of a propagator $P(x,x',m)$ as
\be\label{triple_int}
I_q(n)=A_q\int\limits_{0}^{\infty}\int\limits_{0}^{\infty}\int\limits_{0}^{\infty}dxdydz P(z,y,n)V(y)P(y,x,N-n),
\en
 where $A_q$ is a normalisation constant and $V(x)=e^{-2q\sqrt{a}x}$ with $a=(\avg{z_k^2}-\avg{z_k}^2)/2$. Within the hard wall approximation, the propagator  $P(x,x',m)$ is given by a solution of the classical drift-diffusion equation with an absorbing boundary condition at $x=0$:
\be
 P(x,x',m)=\frac{1}{\sqrt{\pi m}}\big(1-e^{-\frac{4xx'}{m}}\big)e^{-\frac{(x-x'+\gamma m)^2}{m}}\,,
%P(x,x',m)=\frac{e^{-\frac{(x-x'+\gamma m)^2}{m}}}{\sqrt{\pi m}} -\frac{e^{-\frac{(x-x'+\gamma m)^2+4xx'}{m}}}{\sqrt{\pi m}}, 
\en
which satisfies the initial condition $ P(x,x^{\prime},0)=\delta(x-x')$ for $x,x'\ge 0$. A non-zero value  of the drift constant $\gamma=-\avg{z_k}/2\sqrt{a}$ corresponds to the insulating phases, which are distinguished by the sign of $\gamma$, while zero drift corresponds to criticality, and thus $\avg{z_k}=0$ is the condition on the distribution of $t_j$, $t'_j$ that the wavefunction of Eq.~\eqref{wf} is critical.

We briefly summarise the behaviour of the moments away from the critical regime. 
Evaluating the integrals in Eq.~(\ref{triple_int}) for $\gamma\neq 0$, we obtain
\be\label{loc_result}
I_q(n)=\left\{\begin{matrix}\frac{C^q}{q\Gamma(q)}e^{-2\sqrt{a}q\gamma n},\quad \gamma>0\\
\frac{C^q}{q\Gamma(q)}e^{-2\sqrt{a}q|\gamma| (N-n)},\quad \gamma<0\,, \end{matrix}\right.
\en
where $C=1-e^{-2\sqrt{a}\gamma}$, $\Gamma$ is the gamma function, and we assume that $\gamma n\gg \sqrt{n}\gg 1$ for $\gamma>0$,  and $|\gamma| (N-n)\gg \sqrt{N-n}\gg 1$ for $\gamma<0$. Therefore the wavefunction is exponentially localized at the left edge for $\gamma>0$ and at the right edge for $\gamma<0$.
Thus the two  topologically distinct insulating phases in the model can be distinguished by probing the local statistics of the wavefunctions. 
 In terms of the underlying diffusion process, the spatial asymmetry  of the wavefunctions can be interpreted through the interplay of the non-zero drift with the absorbing boundary condition at $x=0$, see the Appendix.

For critical wavefunctions, corresponding to $\gamma=0$, integrating Eq.~(\ref{triple_int})  yields 
 \be\label{mom_bulk}
I_q(n)=\frac{a^{q-1}}{\pi 2^{q-1} q^3\Gamma(q)  \sqrt{n(N-n)}},\quad n,N-n\gg 1.
\en
 The scaling of the moments is $I_q(n)\propto 1/N$ deep in the bulk of the system, where $n\propto N$, which is the scaling law typical for a {\it localized} wavefunction \cite{EM08}. At the same time, the moments get strongly amplified approaching the edges of the system. Exactly at the edge, the moments are given by
\be\label{mom_edge}
I_q(0)=\frac{a^q}{2^qq^2\Gamma(q)\sqrt{\pi a} \sqrt{N}}.
\en 
 The $1/N^{\tau}$-scaling of  $I_q(0)$ with a non-integer exponent $\tau=1/2$ is the scaling law specific for {\it fractal} wavefunctions \cite{EM08}.  The fact that $\tau<1$ is however unusual, as typically amplitudes of the critical wavefunctions are  suppressed rather than enhanced near the boundaries and $\tau>1$ \cite{SGLEMM06}. The inhomogeneous nature of the local moments and the amplification of the critical wavefunction at the edges can be traced to the presence of the localized boundary modes in the insulating phase. In terms of the classical diffusion process it can be traced to the role of the potential $V(x)$, which gives an increased weight to trajectories passing near the absorbing boundary condition at $x=0$. Viewed in this way it is natural that $I_q(0)^2/I_q(N/2)\propto N^{0}$, see the Appendix.

The spatial correlations of the critical wavefunctions can be characterised by the two-point correlation function  $C_{mn}=\avg{|\psi_{2m}\psi_{2n}|^q}$.  Using the same approach as above, we obtain that   $C_{mn}\propto |m-n|^{-3/2}N^{-1}$ in the bulk of the system, and  $C_{0n}\propto n^{-3/2}N^{-1/2}$ at the edge. The power-law scaling of the correlation function is another manifestation of its fractal nature \cite{BF97}.

%%%%%%%%%%%%%%%%% Section %%%%%%%%%%%%%%%%%%%%%%%%%%
%\section{Numerics} 
%%%%%%%%%%%%%%%%% Section %%%%%%%%%%%%%%%%%%%%%%%%%%

%%%%%%%%%%% FIGURE %%%%%%%%% FIGURE %%%%%%%%%%%%%%FIGURE %%%%%%%%% FIGURE %%%%%%%%%%%%%%%%
\begin{figure}[tb]
\centering
\includegraphics[width=0.999\columnwidth]{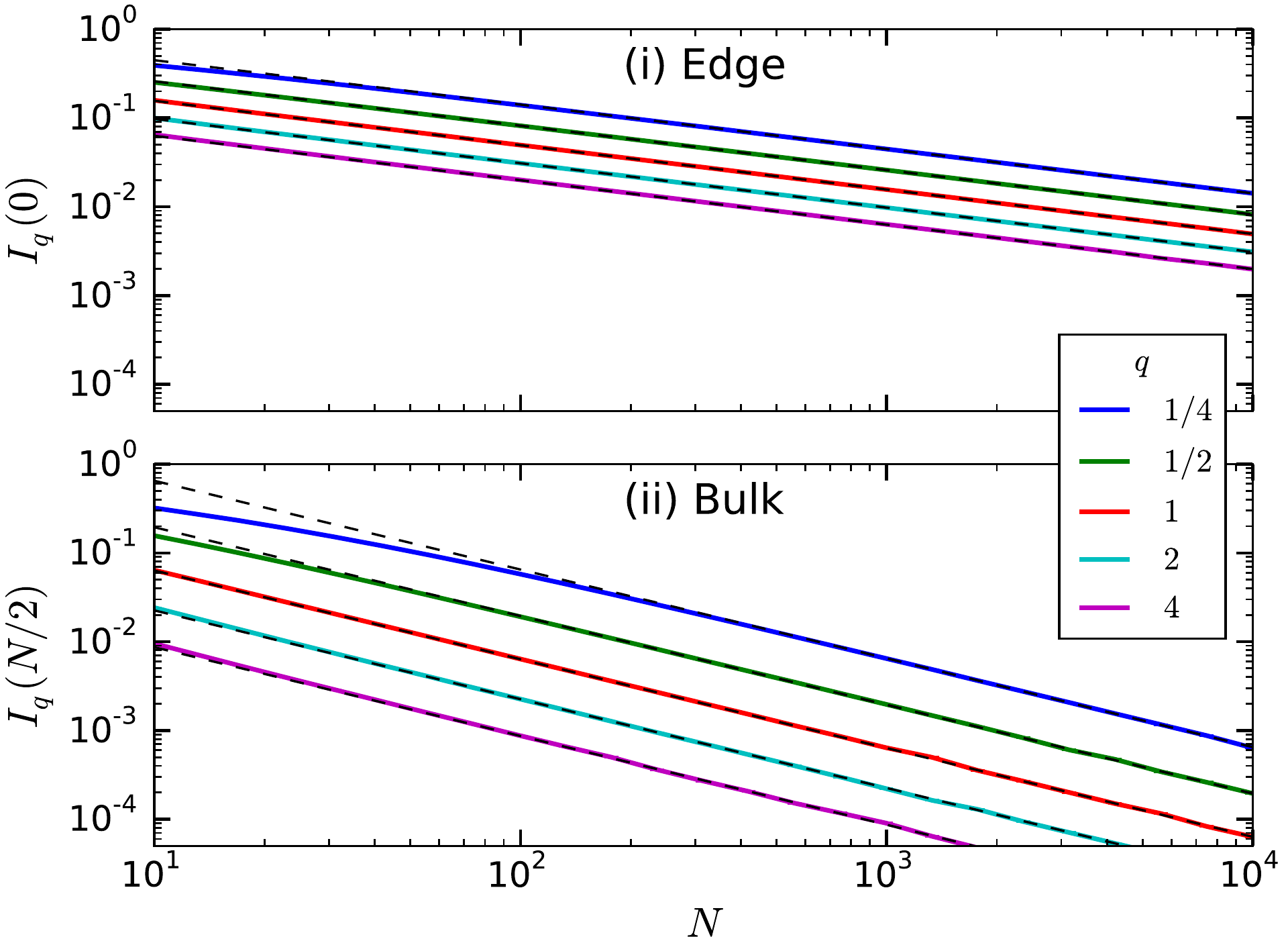}
\caption{\label{fig:edgebulk}
(Color online) The system-size scaling of the local moments $I_q(n)$ of the critical wavefunction of Eq.~\eqref{wf} computed (i) at the edge and (ii) at the centre of the chain. The $q=1/4,1/2,1,2,4$ moments are plotted, and the dashed lines give fits to (i) $1/\sqrt{N}$ scaling at the edge and (ii) $1/N$ scaling in the bulk.
}
\end{figure}
%%%%%%%%%%% FIGURE %%%%%%%%% FIGURE %%%%%%%%%%%%%%FIGURE %%%%%%%%% FIGURE %%%%%%%%%%%%%%%%

%%%%%%%%%%% FIGURE %%%%%%%%% FIGURE %%%%%%%%%%%%%%FIGURE %%%%%%%%% FIGURE %%%%%%%%%%%%%%%%
\begin{figure}[tb]
\centering
\includegraphics[width=0.999\columnwidth]{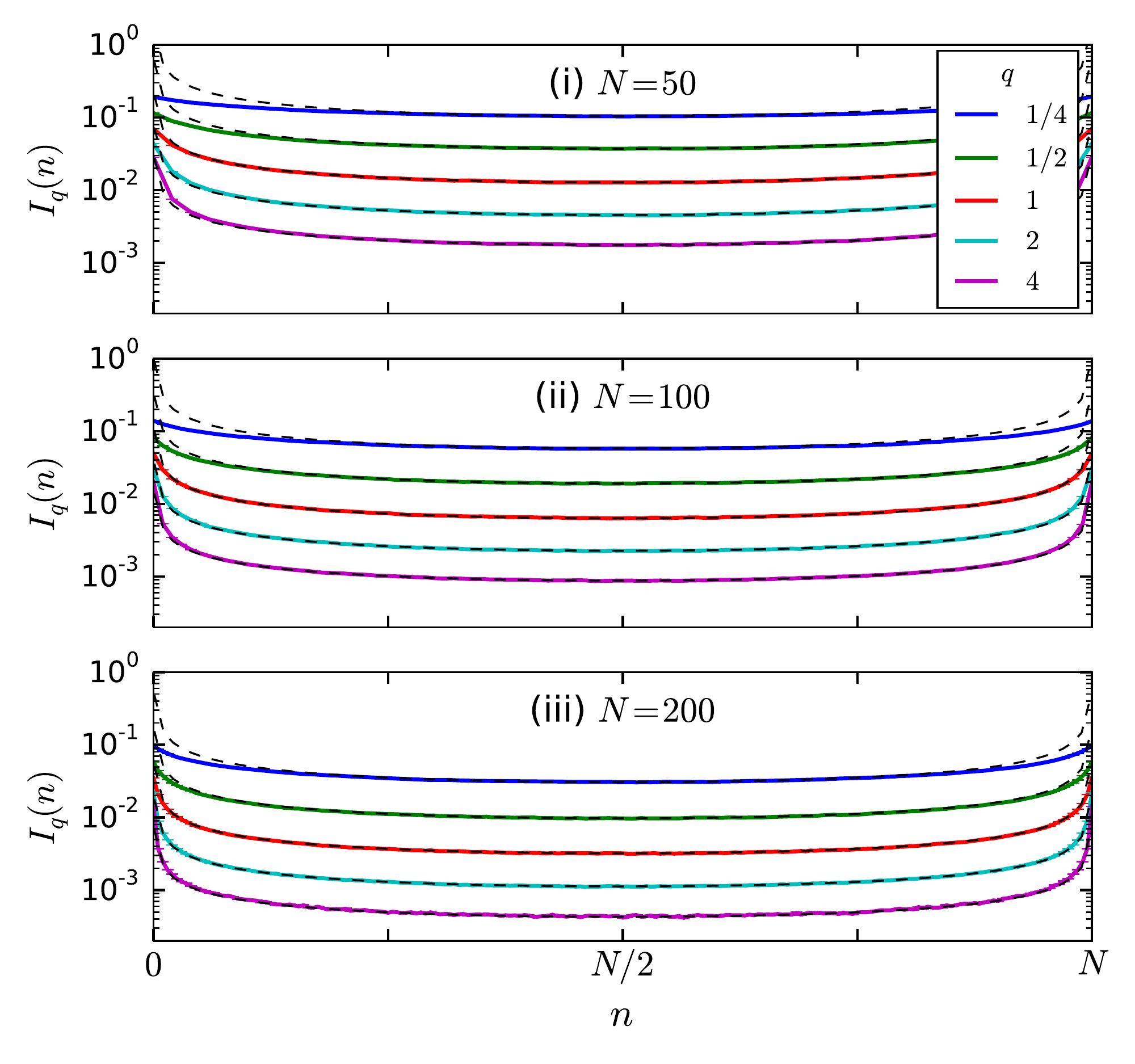}
\caption{\label{fig:inhom1}
(Color online) The spatial profile ($n$ dependence) of the local moments of the critical wavefunction of Eq.~\eqref{wf} with $q=1/4,1/2,1,2,4$ plotted for (i) $N=50$, (ii) $N=100$, (iii) $N=200$. The dashed lines show fits of the data to Eq.~\eqref{mom_bulk} where the normalisation is set by fixing the value at the centre of the chain.
}
\end{figure}
%%%%%%%%%%% FIGURE %%%%%%%%% FIGURE %%%%%%%%%%%%%%FIGURE %%%%%%%%% FIGURE %%%%%%%%%%%%%%%%

\medskip

To investigate the range of validity of the above results, which are derived in the limit of large $N$, we analyse the critical wavefunction numerically. For this we examine a critical point of the Hamiltonian of Eq.~\eqref{poly} given by $t=1$, $t'=0$, $W=2$, $W'=4$, and compute the local moments by averaging the zero-energy wavefunction over $10^6 -10^8$ disorder realisations.
The system-size scaling of the local moments, both directly at the edge and in the centre of the chain, are presented in  Fig.~\ref{fig:edgebulk}. 
It is seen that the edge scaling is more robust than the bulk scaling for small system sizes, while for increasing $N$ an excellent fit to Eq.~\eqref{mom_edge} and  \eqref{mom_bulk}, for the edge and bulk respectively, emerges.

In Fig.~\ref{fig:inhom1}  the spatial profile of the $q=1/4,1/2,1,2,4$ local moments for $N=50,100,200$ are plotted, and compared with a fit to Eq.~\eqref{mom_bulk}.
It is seen that for $q>1$ the formula captures the moments right out to the edge, while for $q<1$ there are deviations near the edges of the chain which get smaller as the system size is increased. Since the moments become uniform in the limit $q\to 0$ one would need diverging system sizes in order to see the predicted form in this regime.

\medskip

%%%%%%%%%%% FIGURE %%%%%%%%% FIGURE %%%%%%%%%%%%%%FIGURE %%%%%%%%% FIGURE %%%%%%%%%%%%%%%%
\begin{figure}[tb]
\centering
\includegraphics[width=0.999\columnwidth]{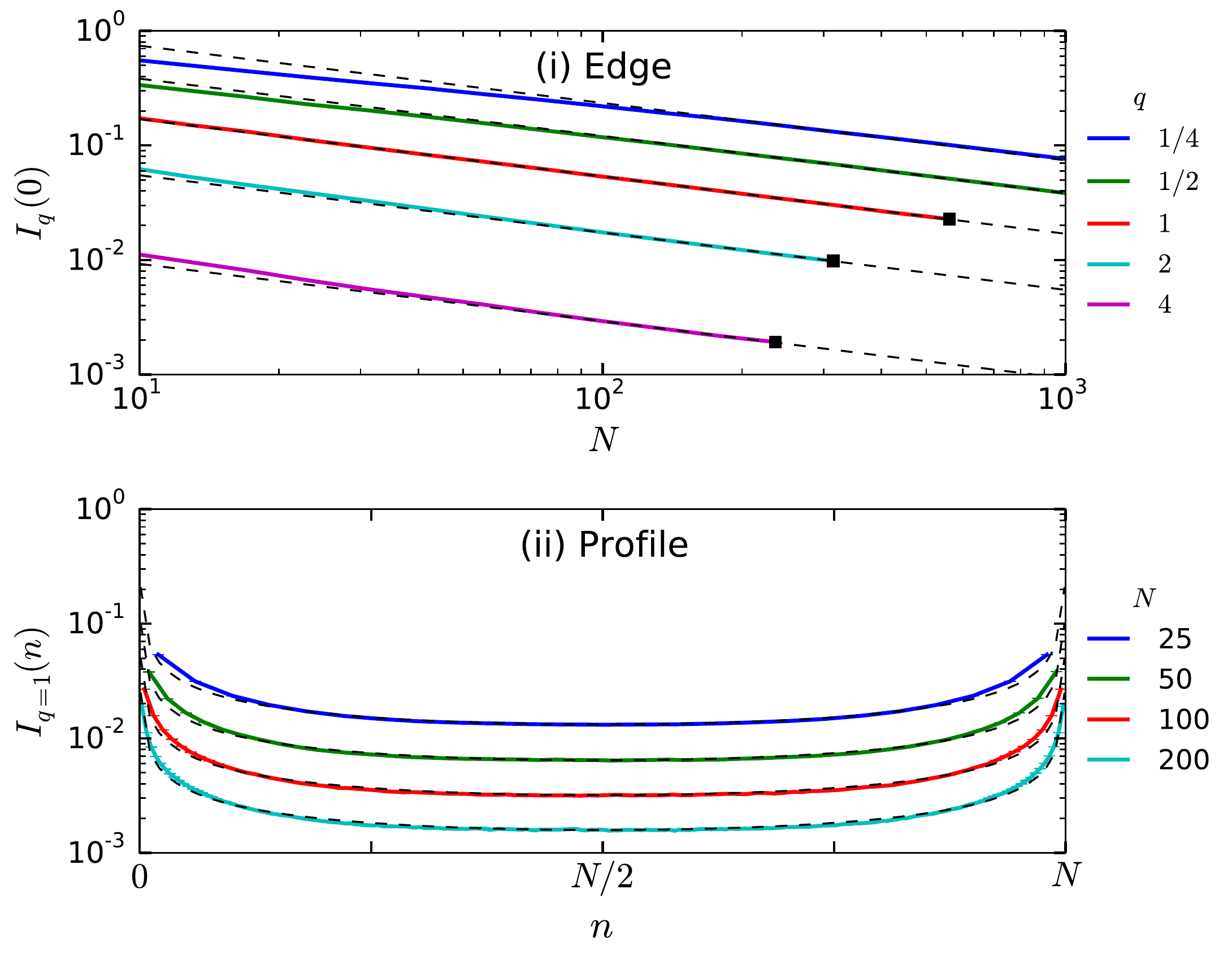}
\caption{\label{fig:edgeinhom2}
(Color online)  Plots of the local moments for the even-length chain, obtained by setting $t_1=0$ in Eq.~\eqref{poly}, with the disorder average taken over wavefunctions whose energy lies within a window about 0, as described in the text. Plot (i) shows the system-size scaling of the edge moments for $q=1/4,1/2,1,2,4$ and the dashed lines give fits to $1/\sqrt{N}$ scaling. The data is terminated at the black squares on the $q=1,2,4$ lines as beyond these system sizes numerical inaccuracy arises due to machine precision. In (ii) the spatial profile of the $q=1$ local moments are plotted for $N=25,50,100,200$ and the dashed lines show fits of the data to Eq.~\eqref{mom_bulk} where the normalisation is set by fixing the value at the centre of the chain.
}
\end{figure}
%%%%%%%%%%% FIGURE %%%%%%%%% FIGURE %%%%%%%%%%%%%%FIGURE %%%%%%%%% FIGURE %%%%%%%%%%%%%%%%

We now perform checks on the generality of our analytic expressions, to support the claim that they are valid at all topological Anderson transition in 1D.  First we consider a dimerised tight-binding chain with an even number of sites, as then there is no wavefunction exactly at $E=0$ for any finite length. To examine the critical properties in this case we take the disorder average $\langle \ldots\rangle$ to run over all wavefunctions in the disorder ensemble whose energy is within a window $[-E_{\rm typ}, E_{\rm typ}]$, where $E_{\rm typ}=\exp(\langle \ln (\min_\a |E_\a|) \rangle)$ with $\{E_\a\}$ the set of eigenenergies for a given realisation. The even length chain is realised by setting $t_1=0$ in Eq.~\eqref{poly}, which decouples the site 0, and it is analysed using exact diagonalisation. 
In Fig.~\ref{fig:edgeinhom2}(i) the system-size scaling of the local moments at the edge of the chain are plotted and compared with the predicted $1/\sqrt{N}$ scaling, while in Fig.~\ref{fig:edgeinhom2}(ii) the spatial profile of the  $q=1$ local moments are presented and compared with the prediction of Eq.~\eqref{mom_bulk}. In both cases a very good fit is seen, which indicates that the results are valid in general and not just to exactly critical wavefunctions.

%%%%%%%%%%%%%%%%% Section %%%%%%%%%%%%%%%%%%%%%%%%%%
%\section{Bogoliubov-de Gennes} 
%%%%%%%%%%%%%%%%% Section %%%%%%%%%%%%%%%%%%%%%%%%%%

%%%%%%%%%%% FIGURE %%%%%%%%% FIGURE %%%%%%%%%%%%%%FIGURE %%%%%%%%% FIGURE %%%%%%%%%%%%%%%%
\begin{figure}[tb]
\centering
\includegraphics[width=0.999\columnwidth]{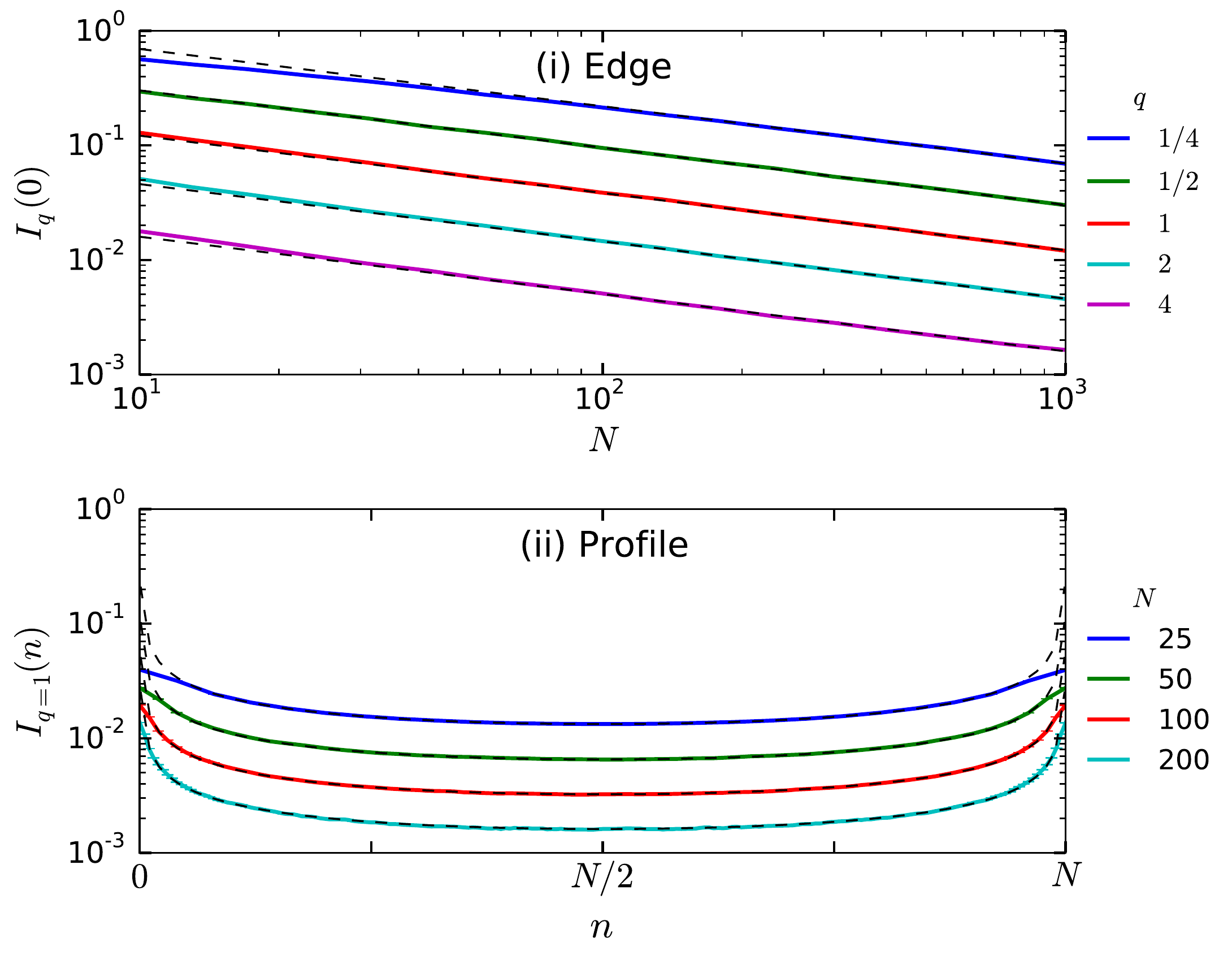}
\caption{\label{fig:edgeinhomD}
(Color online) Plots of the local moments for the Hamiltonian $H_D$ with $t_D=1/4$ and $N_{\rm ch}=5$. The quantities plotted are the same as those in Fig.~\ref{fig:edgeinhom2}, and identical behaviour is observed.
}
\end{figure}
%%%%%%%%%%% FIGURE %%%%%%%%% FIGURE %%%%%%%%%%%%%%FIGURE %%%%%%%%% FIGURE %%%%%%%%%%%%%%%%

The second check on the generality of our results that we perform is to examine a multi-channel model in the Bogoliubov-de Gennes symmetry class D. Specifically we consider the following Hamiltonian
\be\la{HamD}
H_D = \sum_{j=0}^{2N}\big[ -i\,t_D \,\bCd_{j+1} \bC_{j} -i\, \bCd_j B_j \bC_{j}  + {\rm h.c.} \big]\,,
\ee
where $t_D$ is a hopping parameter, $\bC_{j}$ is a column vector $\{\bc_{j,1},\bc_{j,2},\ldots,\bc_{j,N_{\rm ch}}\}$ of $N_{\rm ch}$ fermions at site $j$, $B_j$ is an $N_{\rm ch} \times  N_{\rm ch}$ random matrix satisfying $B_j = - B_j^T$, 
and  $N_{\rm ch}$ is the number of channels.
In practice we set $B_j=A_j-A_j^T$ where $A_j$ are real matrices whose elements  are independent random variables uniformly distributed on the interval $[-0.5,0.5]$. 
The Hamiltonian belongs to symmetry class D when $N_{\rm ch}\ge 2$, as complex conjugation $K$ provides an anti-unitary symmetry that anti-commutes with the Hamiltonian, whereas in the absence of a sublattice symmetry there is generically no anti-unitary symmetry that commutes with the Hamiltonian. The model sits exactly at a topological Anderson transition for any $t_D$, and can be taken into a topological phase by adding a dimerisation to the hopping. 
Indeed, to view the topological properties of the model one must regard it as having a two-site unit cell, and it is for this reason that we set the sum to run to $2N$ in Eq.~\eqref{HamD}.

Again we numerically examine the disorder averaged local moments of the critical wavefunction. For this we take $N_{\rm ch}=5$ and obtain the critical wavefunctions through exact diagonalisation.  
In Fig.~\ref{fig:edgeinhomD} we plot again the system-size scaling of the local moments at the edge of the chain and the spatial profile of the  $q=1$ local moments, as were presented in  Fig.~\ref{fig:edgeinhom2}. We find once more an excellent fit, and  view this as confirmation that our results are generally applicable at topological Anderson transition in 1D.

%\section{Ending} 
\medskip

In summary, from the known exact zero-energy wavefunction of the Su-Schrieffer-Heeger model we have derived analytical expressions for the system-size scaling of the local moments, which are valid at all topological Anderson transitions in 1D. The generality of the expressions follow from the superuniversality of these transitions, and is also evidenced by our numerical simulations of other models, which show excellent agreement despite having distinct features from the original model. The results show that the moments are inhomogneous, with strong amplification towards the edges of the system, and this distinguishes critical wavefunctions at topological Anderson transitions in 1D from all other critical wavefunctions studied so far. It is an interesting open problem whether similar features of critical wavefunctions can be found in higher dimensions.

\bigskip 

\noindent
{\it Acknowledgments } --- A.O. thanks 	
the Max Planck Institute for the Physics of Complex Systems for hospitality. T.C. and A.O. acknowledge support from the London Mathematical Society under Grant No. URB14-37.

\setcounter{equation}{0}
\setcounter{figure}{0}

\renewcommand{\bibnumfmt}[1]{[A#1]}
\renewcommand{\citenumfont}[1]{A#1}
\renewcommand{\thetable}{A\arabic{table}}
\renewcommand{\theequation}{A\arabic{equation}}
\renewcommand{\thefigure}{A\arabic{figure}}

\pagebreak

\appendix

\section{} \la{app}

In this Appendix we provide details of the analytic calculations leading to the results quoted in the main text.

\subsection{Moments of the normalized wavefunctions}
\subsubsection{Moments of the normalized wavefunctions  and classical diffusion}

The explicit form of the wavefunction given by Eq.\ (5) allows us to straightforwardly calculate   its statistical properties, if the normalisation condition is not taken into account. However, for the normalized wavefunction  the problem becomes much more complicated, as the normalisation constant depends on all variables $t_k$, $t'_k$ in a non-trivial way.  In order to overcome this problem we map the calculation to one in Liouville quantum mechanics \cite{ST98a,BF97a}.

First, we define the random variables $x_n=\prod_{k=1}^n\left(\frac{t_k}{t'_k}\right)^2$ and we set $x_0=1$ . Then the disorder averaged local moments of the normalized wavefunction $I_q(n)=\avg{|\psi_{2n}|^{2q}}$ are given by $I_q(n)=\avg{x_n^q/(\sum_{p=0}^Nx_p)^q}$. The denominator of $I_q(n)$ can be represented as
\be
\left(\sum_{p=0}^Nx_p\right)^{-q}=\frac{1}{\Gamma(q)}\int_0^\infty d\alpha \alpha^{q-1}e^{-\alpha\sum_{p=0}^Nx_p}\,,
\en
where $\Gamma$ is the gamma function.
With the help of the above identity we obtain
\be\label{mom}
I_q(n)=\frac{1}{\Gamma(q)}\int_0^\infty d\alpha \alpha^{q-1}e^{-\alpha}J(\alpha),
\en 
where $J(\alpha)=\avg{x_n^q e^{-\alpha\sum_{p=1}^Nx_p} }$ and we used the fact that $x_0=1$. 
The disorder averaging  can be performed using the statistical independence of 
the random variable $z_k=\ln\left(\frac{t_k}{t'_k}\right)^2$:
\begin{widetext}
\be
\J=\int_{-\infty}^{\infty}dz_1\dots dz_N\:P_z(z_1)\dots P_z(z_N)e^{q \sum_{k=1}^n z_k-\alpha \sum_{p=1}^N e^{\sum_{k=1}^pz_k}},
\en
where $P_z(z_k)$ is the distribution function of $z_k$. The structure of the above integrand suggests to introduce new variables
\be
t_p=\sum_{k=1}^pz_k,\quad z_p=t_p-t_{p-1},\: {\rm for}\: p=2,\dots, N\:{\rm and}\: z_1=t_1,
\en
in terms of which $\J$ takes the form
\be
\J=\int_{-\infty}^{\infty}dt_1\dots dt_N\:\prod_{p=1}^NP_z(t_p-t_{p-1})g_p(t_p,\alpha),
\en
where $t_0\equiv 0$, $g_p(t,\alpha)=e^{-\alpha e^t}$ for $p\neq n$ and $g_n(t,\alpha)=e^{qt-\alpha e^t}$. Using the expression for  $P_z(z)$ 
in terms of its Fourier transform  $F(s)=\int_{-\infty}^{\infty}dz e^{isz} P_z(z)$, we obtain
\be
\J=\int_{-\infty}^{\infty}\frac{ds_1}{2\pi}\dots \frac{ds_N}{2\pi}\int_{-\infty}^{\infty}dt_1\dots dt_N\:
\prod_{p=1}^Ne^{-is_p(t_p-t_{p-1})}F(s_p)g_p(t_p,\alpha),
\en
One can notice that 
\be
\left|e^{-is_p(t_p-t_{p-1})}F(s_p)\right|\le |F(s_p)|\le 1=F(0),
\en
hence the main contribution to the integral in the limit $N\to\infty$ comes from $s_p\to 0$. Therefore we can use the Taylor expansion of $F(s)$ about $s=0$. Keeping only the terms up to the quadratic order we can write $F(s)\approx e^{-i b s-as^2}$, where $b=-\avg{z_k}$, 
$a=(\avg{z_k^2}-\avg{z_k}^2)/2$. Then the integrals over $s_p$ become Gaussian and can be evaluated explicitly:
\be
\J=\int_{-\infty}^{\infty}\prod_{p=1}^N\frac{dt_p}{2\sqrt{\pi a}}\:e^{qt_n}
e^{-\left(\sum_{p=1}^N \frac{(t_p-t_{p-1}+b)^2}{4a}+\alpha e^{t_p}\right)}.
\en
This expression can be viewed as a discrete in time path integral for a particle moving in an exponential potential -- Liouville quantum mechanics. One can approximate such a potential by a hard wall potential:
\be
e^{-\alpha e^t}\approx \theta (\z-t),
\en
where $\theta(t)$ is the Heaviside step function, $\alpha e^{\z}=C$ and $C$ is a constant of order $1$. Within this approximation $\J$ is given by
\be\label{path_int}
\J=\int_{-\infty}^{\z}\prod_{p=1}^N\frac{dt_p}{2\sqrt{\pi a}}\:e^{qt_n}
e^{-\sum_{p=1}^N \frac{(t_p-t_{p-1}+b)^2}{4a}}=e^{q\z}
\int_{0}^{\infty}\prod_{p=1}^N\frac{dr_p}{\sqrt{\pi}}\:e^{-2q\sqrt{a}r_n}
e^{-\sum_{p=1}^N (r_p-r_{p-1}-\gamma)^2},
\en
where $r_p=(z-t_p)/2\sqrt{a}$, $r_0=z/2\sqrt{a}$ and $\gamma=b/2\sqrt{a}$. In order to calculate the above integral, we introduce an integral operator $\hat{K}$ defined as
\be
(\K u)(x)=\theta (x)\int_0^{\infty}\frac{dy}{\sqrt{\pi}}e^{-(y-x-\gamma)^2}u(y).
\en
The propagator corresponding to the path integral (\ref{path_int}) is related to $\K$ as
\be
P(x,x^{\prime},m)=\left(\K^m\delta_{x^{\prime}}\right)(x), \quad \delta_{x^{\prime}}(x)\equiv\delta(x-x^{\prime}).
\en 
If $P(x,x^{\prime},m)$ is known, then $\J$ can be found as
\be
\J=e^{q\z}\int_0^{\infty}dx_1\int_0^{\infty}dx_2 P(r_0,x_2,l_2)V(x_2)P(x_2,x_1,l_1),
\en
where $V(x)=e^{-2q\sqrt{a}x}$, $l_1=N-n$, $l_2=n$. In the hard wall approximation $P(x,x^{\prime},m)=0$ for $x<0$, and hence the integration over $\alpha$ in Eq.\ (\ref{mom}) must be restricted to $\alpha<C$. Taking into account that $\alpha^{q-1}e^{q\z}=C^{q}/\alpha$, the expression for the moments reads
\be
I_q(n)=\frac{C^q}{\Gamma(q)}\int_0^Cd\alpha \frac{e^{-\alpha}}{\alpha}\int_0^{\infty}dx_1\int_0^{\infty}dx_2 P(r_0,x_2,l_2)V(x_2)P(x_2,x_1,l_1).
\en
Changing the integration variable $\alpha$ by $\z$ and noting that $e^{-Ce^{-z}}\approx 1$ in the hard wall approximation for $z>0$, we obtain
\be\label{x3_int}
I_q(n)=\frac{2\sqrt{a}C^q}{\Gamma(q)}\int_0^{\infty}dx_1\int_0^{\infty}dx_2 \int_0^{\infty}dx_3 P(x_3,x_2,l_2)V(x_2)P(x_2,x_1,l_1),
\en
where we finally changed $z$ by $x_3=r_0=z/2\sqrt{a}$. This equation is equivalent to Eq.\ (7)  in the main text with $A_q=2\sqrt{a}C^q/\Gamma(q)$.

In order to find the propagator $P(x,x^{\prime},m)$, we notice that, as follows  from its definition,  it satisfies an integral equation
\be\label{int_eq_P}
P(x,x^{\prime},m+1)=\theta (x)\int_0^{\infty}\frac{dy}{\sqrt{\pi}}e^{-(x-y+\gamma)^2}P(y,x^{\prime},m).
\en
Assuming that $P(y,x^{\prime},m)$ is a slowly varying function compared to $e^{-(x-y+\gamma)^2}$ (this assumption is justified by the solution that we obtain, so is self-consistent), one can transform the above integral equation into the drift-diffusion equation with the absorbing boundary condition at $x=0$. The solution of the drift-diffusion equation obtained by the standard eigenfunction method reads
\be\label{diff_eq_sol}
 P(x,x^{\prime},m)=\frac{1}{\sqrt{\pi m}}\left(e^{-\frac{(x-x'+\gamma m)^2}{m}}-e^{-\frac{(x-x'+\gamma m)^2+4xx'}{m}}\right).
\en
One can check by direct calculation that the above expression for $P(x,x^{\prime},m)$ solves the integral equation (\ref{int_eq_P}) provided that $m,x \gg 1$, i.e. it works well for large times and far away from the boundary $x=0$. Eq.\ (\ref{x3_int}) and Eq.\ (\ref{diff_eq_sol}) give a complete description of the local moments of the wavefunctions in terms of classical diffusion. 

%%%%%%%%%%%%%%%%%%%%%%%%%%%%%%%%%%%%%%%%%%%%%%%%

\subsubsection{Localized wavefunctions}

The insulating phase is characterized by a non-zero drift constant $\gamma\neq 0$.   Substituting  Eq.\ (\ref{diff_eq_sol}) into Eq.\ (\ref{x3_int}) and integrating over $x_1$ and $x_3$ we obtain
\be
I_q(n)=\frac{\sqrt{a}C^q}{2\Gamma(q)}\int_0^{\infty}dx
\left(1+\erf{u_+(x)}-e^{-4x\gamma}\erfc{u_-(x)}\right)
\left(1+\erf{v_-(x)}-e^{4x\gamma}\erfc{v_+(x)}\right)
 e^{-2q\sqrt{a}x},
\en
where $u_{\pm}(x)=(x\pm \gamma l_1)/{\sqrt{l_1}}$, $v_{\pm}(x)=(x\pm\gamma l_2)/{\sqrt{l_2}}$, $\erf{x}$ is the error function and $\erfc{x}=1-\erf{x}$. To evaluate this integral we may use the following approximation for $\erf{x}$
\be
\erf{x}\approx 2\theta(x)-1,
\en
which is justified in the above integral provided that $\gamma l_i\gg\sqrt{l_i}\gg 1$. This approximation yields the following result for $I_q(n)$ 
\be\label{loc_result2}
I_q(n)=\begin{cases} \frac{C^q}{q\Gamma(q)}e^{-2\sqrt{a}q\gamma n}\,, &\gamma>0 \\ 
\frac{C^q}{q\Gamma(q)}e^{-2\sqrt{a}q|\gamma| (N-n)}\,, &\gamma<0\,. \end{cases}
\en
The positive $\gamma$ case corresponds to the wave function localized at the left edge, while the negative $\gamma$ case corresponds to the wave function localized at the right edge. Thus the two different topological phases in our model can be distinguished by probing the local statistics of the wavefunctions.

In order to determine $C$ we use the normalization condition  $\sum_{n=0}^N I_1(n)=1$, which gives
\be
C=1-e^{-2\sqrt{a}\gamma}.
\en
The spatial asymmetry of the result can be naturally understood in terms of the underlying diffusion process. Let us consider for simplicity $\gamma>0$ and two sub-cases $n=0$ and $n=N$.  For positive $\gamma$, the propagator $P(x,x',m)$ describes trajectories with a drift to the left (the mean value of $x$ decreases in "time" $m$), as it follows from Eq.\ (\ref{diff_eq_sol}). Now consider the wavefunction at $n=0$, the moments are given by 
\be\label{x3_int_edge_loc}
I_q(0)=\frac{2\sqrt{a}C^q}{\Gamma(q)}\int_0^{\infty}dx_1\int_0^{\infty}dx_2 V(x_2)P(x_2,x_1,N).
\en
The propagator $P(x_2,x_1,N)$ describes trajectories of classical particles, which start from $x=x_1$ at time $m=0$ and propagate to $x=x_2$ at $m=N$. The contribution of each trajectory is weighted with  an exponential function $V(x_2)=e^{-2q\sqrt{a}x_2}$ at the final point. For this reason the main contribution to the results comes from the trajectories with $x_2$ close to zero. Such trajectories can be found and, due to the drift term, the most probable starting point for them is $x_1=\gamma N$. The contribution of such trajectories yields the result of order of one. 

If we now consider the opposite edge  $n=N$, then
\be
I_q(N)=\frac{2\sqrt{a}C^q}{\Gamma(q)}\int_0^{\infty}dx_1\int_0^{\infty}dx_2 P(x_2,x_1,N) V(x_1).
\en
The weight function  $V(x_1)=e^{-2q\sqrt{a}x_1}$ is associated with the starting point, which forces the initial coordinate $x_1$ to be close to zero. On the other hand, the absorbing boundary condition at $x=0$, together with the presence of the drift term which tends to decrease the position of a particle in time,   lead to the exponential small survival probability of such trajectories and to the exponentially small final result.

%%%%%%%%%%%%%%%%%%%%%%%%%%%%%%%%%%%%%%%%%%%%%%%%

\subsubsection{Critical wavefunctions}

The critical phase is characterised by a vanishing drift constant $\gamma=0$. In this case integration over $x_3$ and $x_1$ in Eq.\ (\ref{x3_int}) leads to the following result:
\be
I_q(n)=\frac{2\sqrt{a}C^q}{\Gamma(q)}\int_0^{\infty}dx\: \erf{\frac{x}{\sqrt{l_2}}} \erf{\frac{x}{\sqrt{l_1}}}e^{-2q\sqrt{a}x}.
\en
Recalling that $l_1=N-n$, $l_2=n$ and assuming that the site $n$ is in the bulk of the system, we are allowed to replace two error functions in the thermodynamic limit $N\to \infty$  by their asymptotic expressions for $l_1, l_2\to \infty$
\be
 \erf{\frac{x}{\sqrt{l}}}=\frac{2x}{\sqrt{\pi l }}+O\left(l^{-3/2}\right).
\en
Then the remaining integration over $x$ yields
\be\label{mom_C}
I_q(n)=\frac{2C^q}{q^3\Gamma(q)\pi a \sqrt{n(N-n)}}.
\en
In order to fix the value of the constant $C$ we use again the normalization condition $\sum_{n=0}^N I_1(n)=1$, which gives $C=a/2$.

If $n=\beta N$ with $N$-independent constant $\beta$, then  $I_q(n)\propto 1/N$, which is the scaling law typical for a {\it localized} wavefunction. 

However, if we are interested in the scaling of the local moments on a site which is close to the boundary, the result will be very different. Indeed, let us consider $n=0$. Then using Eq.\ (\ref{x3_int_edge_loc}) and the same steps as above, we obtain
\be\label{mom_edge2}
I_q(0)=\frac{C^q}{q^2\Gamma(q)\sqrt{\pi a} \sqrt{N}}.
\en 
Thus $I_q(0)\propto 1/\sqrt{N}$, which is the scaling law typical for a {\it fractal} wave function. The same scaling is valid for the opposite edge $n=N$.

The two different results for the scaling of the moments at the edge and in the bulk of the system can be traced to the presence of the localized boundary modes in the insulating phase. It can be also understood in terms of the classical diffusion process as follows. The propagator $P(x_2,x_1,N)$ describes trajectories of classical particles, which start from $x=x_1$ at time $m=0$ and propagate to $x=x_2$ at $m=N$. In Eq.\ (\ref{x3_int_edge_loc}) the contribution of each trajectory is weighted with  an exponential function $V(x_2)=e^{-2q\sqrt{a}x_2}$ at the final point. For this reason the main contribution to the result comes from the trajectories with $x_2$ close to zero. Thus  Eq.\ (\ref{x3_int_edge_loc}) gives a probability of getting from any starting  point $x_1$ to an  $x_2$ in the vicinity of zero  in time $N$, and according to our calculations this scales as $1/\sqrt{N}$. Similarly, Eq.\ (\ref{x3_int}) for $n=N/2$ counts the contributions from the trajectories starting at any point $x_1$ reaching  an  $x_2$ in the vicinity of zero in time $N/2$, and then reaching an arbitrary final point  $x_3$ in time $N/2$. The probability for the first part of the process scales as $1/\sqrt{N}$, but the same is true for the second part of the process due to the time reversal invariance of the problem. Hence the total probability must scale as $(1/\sqrt{N})^2=1/N$.
 
%%%%%%%%%%%%%%%%%%%%%%%%%%%%%%%%%%%%%%%%%%%%%%%%

\subsection{Two-point correlation function}

Spatial correlations of the wavefunctions can be determined through the two-point correlation function
\be
C_{mn}=\avg{|\psi_m\psi_n|^q}.
\en 
Performing all the steps which led us to Eq.\ (\ref{x3_int}) we obtain
\be\label{corr_bulk}
C_{mn}=\frac{2\sqrt{a}C^q}{\Gamma(q)}\int_0^{\infty}dx_1\int_0^{\infty}dx_2 \int_0^{\infty}dx_3\int_0^{\infty}dx_4  P(x_4,x_3,l_3)U(x_3)P(x_3,x_2,l_2)U(x_2)P(x_2,x_1,l_1),
\en
where  $U(x)=e^{-q\sqrt{a}x}$, $l_1=N-n$, $l_2=n-m$, $l_3=m$. Integration over $x_1$ and $x_4$, and expansion of the resulting error functions gives
\be\label{cor_int_x}
C_{mn}=\frac{8\sqrt{a}C^q}{\pi\Gamma(q)\sqrt{l_1l_3}}\int_0^{\infty}dx_2 \int_0^{\infty}dx_3  x_3e^{-q\sqrt{a} x_3} P(x_3,x_2,l_2) 
x_2e^{-q\sqrt{a} x_2} .
\en 
The above integral can be computed explicitly in terms of the error functions. However,  in order to find how it scales with $|m-n|$,  it is more convenient to write it in the momentum space using the complete set of standing waves \cite{BF97a}
\be
\langle z|k\rangle=\sqrt{\frac{2}{\pi}}\sin(kz), \quad k>0.
\en 
Writing the propagator in the momentum representation
\be
P(x_3,x_2,l_2) =\langle x_3|\hat{P}(l_2)|x_2\rangle=\int_0^\infty dk \int_0^\infty dk' \langle x_3|k\rangle \langle k|\hat{P}(l_2)|k'\rangle \langle k'|x_2\rangle,
\en
calculating the matrix elements
\be
\langle k|\hat{P}(l_2)|k'\rangle=\frac{2}{\pi}\int_0^\infty dx \int_0^\infty dy \sin(kx)\sin(k'y)P(x,y,l_2)= e^{-l_2k^2/4}\delta(k-k'),
\en
and substituting two equations above into Eq.\ (\ref{cor_int_x}), we find
\be\label{corr_k_int}
C_{mn}=\frac{8\sqrt{a}C^q}{\pi\Gamma(q)\sqrt{l_1l_3}}\int_0^{\infty}dk e^{-l_2k^2/4}\left( \int_0^{\infty}dx \langle x|k\rangle xe^{-q\sqrt{a} x} \right)^2
\en
Evaluating the integral over $x$ and scaling the variable $k\to q\sqrt{a} k$ we obtain the final result
\be
C_{mn}=\frac{64 C^q}{\pi^2 a q^3\Gamma(q)\sqrt{l_1l_3}}\int_0^{\infty}dk \frac{k^2e^{-l_2q^2 a k^2/4}}{(k^2+1)^4},
\en
which coincides up to a constant with the result by Balents and Fisher \cite{BF97a}.  For $l_2=0$, the integral over $k$ yields $\pi/32$ \footnote{ In Ref.\ \cite{BF97a} the value of this integral was erroneously assumed to be equal to $\pi/16$.} and hence we reproduce the expression for the moments (\ref{mom_C}). For $l_2\gg 1$ and $m$ and $n$ of order of $N\gg 1$ we find
\be
C_{mn}\propto |m-n|^{-3/2}N^{-1},
\en
showing that the critical wavefunctions exhibit  power law correlations in the bulk of the system.

If one of the points is located at the edge of the system $m=0$, then the above derivation should be modified. Eq.\ (\ref{corr_bulk}) must be replaced by
\be
C_{0n}=\frac{2\sqrt{a}C^q}{\Gamma(q)}\int_0^{\infty}dx_1\int_0^{\infty}dx_2 \int_0^{\infty}dx_3 
U(x_3)P(x_3,x_2,l_2)U(x_2)P(x_2,x_1,l_1),
\en
where $l_1=N-n$, $l_2=n$. Performing all the  steps similar to the ones, which led to Eq.\ (\ref{corr_k_int}), we find
\be
C_{0n}=\frac{4\sqrt{a}C^q}{\sqrt{\pi}\Gamma(q)\sqrt{l_1}}\int_0^{\infty}dk e^{-l_2k^2/4} \int_0^{\infty}dx_3 \langle x_3|k\rangle e^{-q\sqrt{a} x_3}  \int_0^{\infty}dx_2 \langle x_2|k\rangle x_2e^{-q\sqrt{a} x_2}.
\en
Computing the integrals over $x_2$ and $x_3$ we arrive at the final result
\be
C_{0n}=\frac{16 C^q}{\pi^{3/2} \sqrt{a} q^2\Gamma(q)\sqrt{l_1}}\int_0^{\infty}dk \frac{k^2e^{-l_2q^2 a k^2/4}}{(k^2+1)^3}.
\en
For $l_2=0$, the integral over $k$ gives $\pi/16$ and we reproduce Eq.\ (\ref{mom_edge2}). For $l_2\gg 1$ and $N-n$ of order of $N$ we find
\be
C_{0n}\propto n^{-3/2}N^{-1/2},
\en
which has the same scaling with $m$ as in the bulk, but a different  scaling with $N$.

 Finally,  for the both points located at the opposite edges an analogue of  Eq.\ (\ref{corr_bulk}) reads
\be
C_{0N}=\frac{2\sqrt{a}C^q}{\Gamma(q)}\int_0^{\infty}dx_2 \int_0^{\infty}dx_3 
U(x_3)P(x_3,x_2,N)U(x_2),
\en
which can be transformed to the momentum representation as
\be
C_{0N}=\frac{2\sqrt{a}C^q}{\Gamma(q)}\int_0^{\infty}dk e^{-Nk^2/4}\left( \int_0^{\infty}dx \langle x|k\rangle e^{-q\sqrt{a} x}\right)^2.
\en
Calculating the integral over $x$, we obtain
\be
C_{0N}=\frac{4C^q}{\pi q\Gamma(q)}\int_0^{\infty}dk \frac{k^2 e^{-Nq^2 a k^2/4}}{(k^2+1)^2},
\en
which implies that $C_{0N}\propto N^{-3/2}$.

\end{widetext}

\end{document}